\documentstyle[ltwol]{article}

\input{psfig}

\def\NPB{{\em Nucl. Phys.} B~}
\def\PLB{{\em Phys. Lett.}  B~}
\def\PRL{{\em Phys. Rev. Lett.}~}
\def\PRD{{\em Phys. Rev.} D~}
\def\ZPC{{\em Z. Phys.} C~}

\def\be{\begin{eqnarray}}
\def\en{\end{eqnarray}}
\def\ra{\rangle}
\def\la{\langle}
\def\non{\nonumber}
\def\bi{\bibitem}
\def\nc{N_c^{\rm eff}}
\def\ov{\overline}
\def\lsim{ {\ \lower-1.2pt\vbox{\hbox{\rlap{$<$}\lower5pt\vbox{\hbox{$\sim$}
}}}\ } }
\def\gsim{ {\ \lower-1.2pt\vbox{\hbox{\rlap{$>$}\lower5pt\vbox{\hbox{$\sim$}
}}}\ } }

\input psfig

\begin{document}

\title{NONFACTORIZABLE EFFECTS IN CHARMLESS $B$ DECAYS AND $B$ 
MESON LIFETIMES}

\author{HAI-YANG CHENG}

\address{Institute of Physics, Academia Sinica, Taipei, Taiwan 115, Republic
of China \\E-mail: phcheng@ccvax.sinica.edu.tw}


\twocolumn[\maketitle\abstracts{Status of nonfactorizable effects in hadronic 
charmless $B$ decays is reviewed. Implications of new CLEO measurements on
$B^0\to\pi^+\pi^-$ and $B\to\eta' K$ are discussed. Nonfactorizable effects
due to color ocet 4-quark operators are calculated using renormalization
group improved QCD sum rules. The resultant $B$-meson lifetime
ratio $\tau(B^-)/\tau(B_d)$ agrees with experiment.
}] 

\section{Generalized Factorization}
  The nonleptonic two-body decays of mesons are conventionally evaluated under
the factorization hypothesis. In the factorization approach, the decay 
amplitude is expressed in terms of factorizable hadronic matrix elements 
multiplied by some combinations of Wilson coefficient functions. To be more
specific,  the factorization hypothesis assumes that the 3-body hadronic
matrix element $\la h_1h_2|O|M\ra$ for the decay $M\to h_1h_2$ is 
approximated as th product of two matrix elements $\la h_1|J_{1\mu}|0\ra$
and $\la h_2|J_2^\mu|M\ra$. However, it is known that
this approach of naive factorization fails
to describe the decays proceeding through the (class-I) color-suppressed 
internal $W$-emission diagrams, though it is at work for decay modes 
dominated by (class-II) external $W$-emission diagrams. This implies that it is
necessary to take into account nonfactorizable contributions to the 
decay amplitude in order to render
the color suppression of internal $W$-emission ineffective.

Because there is only one single form factor
involved in the class-I or class-II decay amplitude of $B\to PP,~PV$ decays, 
the effects of nonfactorization can be lumped into the
effective parameters $a_1$ and $a_2$: \cite{Cheng94}
\be
a_{1,2}^{\rm eff} = c_{1,2}^{\rm eff}+c_{2,1}^{\rm eff}\left({1\over 
N_c}+\chi_{1,2}\right),
\en
where $\chi_i$ are nonfactorizable terms and receive main contributions from
color-octet current operators. 
Since $|c_1/c_2|\gg 1$, it is evident from 
Eq.~(1) that even a small amount of nonfactorizable contributions will have a
significant effect on the color-suppressed class-II amplitude.
If $\chi_{1,2}$ are universal (i.e. process independent) in
charm or bottom decays, then we have a generalized factorization scheme
in which the decay amplitude is expressed in terms of factorizable 
contributions multiplied by the universal effective parameters
$a_{1,2}^{\rm eff}$. For $B\to VV$ decays, this new factorization implies
that nonfactorizable terms contribute in equal weight to all partial wave
amplitudes so that $a_{1,2}^{\rm eff}$ {\it can} be defined.
It should be stressed that, contrary to the
naive one, the improved factorization does incorporate nonfactorizable
effects in a process independent form.
Phenomenological analyses
of two-body decay data of $D$ and $B$ mesons indicate that while
the generalized factorization hypothesis in general works reasonably well, 
the effective parameters $a_{1,2}^{\rm eff}$ do show some variation from
channel to channel, especially for the weak decays of charmed mesons
 \cite{Cheng94,Kamal96,Cheng96}.
An eminent feature emerged from the data analysis is that $a_2^{\rm eff}$ 
is negative in charm decay, whereas it becomes positive in the two-body decays
of the $B$ meson \cite{Cheng94,CT95,Neubert}:
\be
a_2^{\rm eff}(B\to D\pi)& \sim& 0.20-0.28\,,   \non \\ 
\chi_2(B\to D\pi) &\sim & 0.12-0.19\,.
\en
Phenomenologically, it is often to treat the number of colors $N_c$ as
a free parameter to model the nonfactorizable contribution to hadronic
matrix elements and its value can be extracted from the data of two-body
nonleptonic decays. Theoretically, this amounts to
defining an effective number of colors $\nc$, called $1/\xi$ in \cite{BSW87},
by $1/N_c^{\rm eff}\equiv (1/N_c)+\chi$. It is clear from (2) that
\be
N_c^{\rm eff}(B\to D\pi)=1.8-2.2\approx 2\,.
\en

\section{Nonfactorizable Effects in Hadronic Charmless $B$ Decays}

What are the nonfactorizable effects in hadronic charmless $B$ decays 
\cite{CLEO}?
We note that the effective Wilson coefficients appear in the factorizable 
decay amplitudes in the combinations $a_{2i}=
{c}_{2i}^{\rm eff}+{1\over N_c}{c}_{2i-1}^{\rm eff}$ and $a_{2i-1}=
{c}_{2i-1}^{\rm eff}+{1\over N_c}{c}^{\rm eff}_{2i}$ $(i=1,\cdots,5)$. 
As discussed in Sec.~1.
nonfactorizable effects in the 
decay amplitudes of $B\to PP,~VP$ can be absorbed into the parameters
$a_i^{\rm eff}$. This amounts to replacing $N_c$ in $a^{\rm eff}_i$
by $(N_c^{\rm eff})_i$. Explicitly,
\be
a_{2i}^{\rm eff} &=& {c}_{2i}^{\rm eff}+{1\over (N_c^{\rm eff})_{2i}}
{c}_{2i-1}^{\rm eff}, \non \\ 
a_{2i-1}^{\rm eff} &=&
{c}_{2i-1}^{\rm eff}+{1\over (N_c^{\rm eff})_{2i-1}}{c}^{\rm eff}_{2i}.
\en
It is customary to assume in the literature that $(N_c^{\rm eff})_1
\approx (N_c^{\rm eff})_2\cdots\approx (N_c^{\rm eff})_{10}$;
that is, the nonfactorizable
term is usually assumed to behave in the same way in tree and penguin 
decay amplitudes. A closer investigation shows 
that this is not the case. We have argued in \cite{CT98} that
nonfactorizable effects in the matrix
elements of $(V-A)(V+A)$ operators are different from that of 
$(V-A)(V-A)$ operators. One reason is that
the Fierz transformation of the $(V-A)(V+A)$ operators $O_{5,6,7,8}$
is quite different from that of $(V-A)(V-A)$ operators $O_{1,2,3,4}$
and $O_{9,10}$. Hence, we will advocate that
\be
N_c^{\rm eff}(LL) &\equiv&
\left(N_c^{\rm eff}\right)_{1,2,3,4,9,10}\,, \non \\
 N_c^{\rm eff}(LR) &\equiv &
\left(N_c^{\rm eff}\right)_{5,6,7,8}\,,
\en
and that $N_c^{\rm eff}(LR)\neq N_c^{\rm eff}(LL)$. In principle, 
$N_c^{\rm eff}$ can vary from channel to channel, as in the case of charm
decay. However, in the energetic two-body $B$ decays, $\nc$
is expected to be process insensitive as supported by data 
\cite{Neubert}. 

\subsection{Nonfactorizable effects in spectator amplitudes}
To study $\nc(LL)$ in spectator amplitudes, 
we focus on the class-III decay modes sensitive to
the interference between external and internal $W$-emission amplitudes. 
Good examples are the class-III modes: $B^\pm\to \omega\pi^\pm,~\pi^0
\pi^\pm,~\eta\pi^\pm,~\pi^0\rho^\pm,\cdots$, etc. Considering $B^\pm\to
\omega\pi^\pm$, we find that
the branching ratio is sensitive to $1/\nc$ and has the 
lowest value of order $2\times 10^{-6}$ at $\nc=\infty$ and then increases 
with $1/\nc$. The 1997 CLEO measurement yields \cite{CLEOomega1}
\be
{\cal B}(B^\pm\to\omega\pi^\pm)=\left(1.1^{+0.6}_{-0.5}\pm 0.2\right)
\times 10^{-5}.
\en
Consequently, $1/\nc>0.35$ is preferred by the data \cite{CT98}. 
Because this decay 
is dominated by tree amplitudes, this in turn implies that $\nc(LL)<2.9$.
If the value of $\nc(LL)$ is fixed to be 2, the branching ratio 
for positive $\rho$, which is preferred by the current analysis 
\cite{Parodi}, will be of order $(0.9-1.0)\times 10^{-5}$, which is 
very close to the central value of the measured one. Unfortunately,
the significance of $B^\pm\to\omega\pi^\pm$ is reduced in the
recent CLEO analysis and only an upper limit is quoted \cite{CLEOomega2}:
${\cal B}(B^\pm\to\pi^\pm\omega)<2.3\times 10^{-5}$.
Nevertheless, the central value of ${\cal B}(B^\pm\to
\pi^\pm\omega)$ remains about the same as (6). The fact that $\nc(LL)\sim 2$ 
is preferred in charmless two-body decays of the $B$ meson 
is consistent with the nonfactorizable term extracted from $B\to (D,D^*) 
(\pi,\rho)$ decays: $\nc(B\to D\pi)\approx 2$. Since the
energy release in the energetic two-body decays $B\to\omega\pi$, $B\to D\pi$
is of the same order of magnitude, it is thus expected that $\nc(LL)|_{B
\to\omega\pi}\approx 2$.

 In analogue to the decays $B\to D^{(*)}(\pi,\rho)$, the interference effect of
spectator amplitudes in class-III charmless $B$ decay can be tested 
by measuring the ratios:
\be  
R_1 &\equiv& 2\,{{\cal B}(B^-\to\pi^-\pi^0)\over {\cal B}(\ov B^0\to \pi^-
\pi^+)},   \non \\
R_2 &\equiv& 2\,{{\cal B}(B^-\to\rho^-\pi^0)\over {\cal B}(\ov 
B^0\to \rho^-\pi^+)},  \non \\ 
R_3 &\equiv& 2\,{{\cal B}(B^-\to\pi^-\rho^0)\over {\cal B}(\ov B^0\to 
\pi^-\rho^+)}.
\en
The ratios $R_i$ are greater (less) than unity when the 
interference is constructive (destructive). 
Hence, a measurement of $R_i$ (in particular $R_3$) \cite{CT98}, which has 
the advantage of being independent of
the Wolfenstein parameters $\rho$ and $\eta$, will constitute a very 
useful test on the effective number of colors $\nc(LL)$. 

During this conference, CLEO has reported the updated limits
on ${B}^0\to\pi^+\pi^-$ and $B^-\to\pi^-\pi^0$: \cite{Roy}
\be
&& {\cal B}({B}^0\to\pi^+\pi^-)<0.84 \times 10^{-5}, \non \\
&& {\cal B}(B^-\to\pi^-\pi^0)<1.6\times 10^{-5}.
\en
In particular, the limit on $B^0\to\pi^+\pi^-$ is improved by a factor of 2. 
It 
appears that this decay provides a stringent constraint on the form factor
$F_0^{B\pi}$. Irrespective of the values of $\nc$, the predicted branching
ratio for $B^0\to\pi^+\pi^-$ will easily exceed the current limit if 
$F_0^{B\pi}
(0)\gsim 0.30$. Note that the decay rate of $B^0\to\pi^+\pi^-$ increases 
slightly with $\nc(LL)$ as it is dominated by the tree coefficient $a_1$. 
For $F_0^{B\pi}(0)=0.30$, we find $\nc(LL)\lsim 0.20$\,.

\subsection{Nonfactorizable effects in penguin amplitudes}
The penguin amplitude of the class-VI mode $B\to \phi K$ is proportional to
the QCD penguin coefficients $(a_3+a_4+a_5)$ and hence sensitive to the 
variation of $\nc(LR)$ since $a_4$ is $\nc$-stable, but $a_3$ and $a_5$ are
$\nc$-sensitive. Neglecting $W$-annihilation and space-like penguin diagrams, 
we find \cite{CT98} that $\nc(LR)=2$
is evidently excluded from the present CLEO upper limit \cite{CLEOomega2}
\be \label{phiK}
{\cal B}(B^\pm\to\phi K^\pm)< 0.5\times 10^{-5},
\en
and that $1/\nc(LR)<0.23$ or $\nc(LR)>4.3\,$.
A similar observation was also made in \cite{Deshpande2}. 
The branching ratio of $B\to\phi K^*$, the average of $\phi K^{*-}$ and
$\phi K^{*0}$ modes, is also measured recently by CLEO with the result
\cite{CLEOomega2}
\be    \label{phiK*}
{\cal B}(B\to\phi K^*) 
=\left(1.1^{+0.6}_{-0.5}\pm 0.2\right)\times 10^{-5}.
\en
We find that the allowed region for $\nc(LR)$ is $4\gsim \nc(LR)\gsim 1.4$.
This is in contradiction to the constraint
$\nc(LR)>4.3$ derived from $B^\pm\to\phi K^\pm$. In fact,  
the factorization approach predicts that
$\Gamma(B\to\phi K^*)\approx\Gamma(B\to\phi K)$ 
when the $W$-annihilation type of contributions is neglected. The current
CLEO measurements (\ref{phiK}) and (\ref{phiK*}) are obviously not consistent 
with the
prediction based on factorization. One possibility is that
generalized factorization is not applicable to $B\to VV$.
Therefore, the discrepancy between ${\cal B}(B\to\phi K)$ and 
${\cal B}(B\to\phi K^*)$ will measure
the degree of deviation from the generalized factorization that has been
applied to $B\to\phi K^*$. It is also possible that the absence 
of $B\to\phi K$ events is a downward statistical fluctuation. At any rate, 
in order to clarify this issue and to pin down the effective number of 
colors $\nc(LR)$, we urgently need 
measurements of $B\to\phi K$ and $B\to\phi K^*$, especially the neutral 
modes, with sufficient accuracy. 

\subsection{$B\to\eta' K$ decays}

The published CLEO results \cite{Behrens1} on the decay $B\to\eta' K$ 
\be
{\cal B}(B^\pm\to\eta' K^\pm) &=& \left(6.5^{+1.5}_{-1.4}\pm 0.9\right)\times
10^{-5}, \non \\
 {\cal B}(B^0\to\eta' K^0) &=& \left(4.7^{+2.7}_{-2.0}\pm 0.9
\right)\times 10^{-5},
\en
are several times larger than previous theoretical predictions
\cite{Chau1,Kramer,Du} in the range of $(1-2)\times 10^{-5}$.
It was pointed out last year by several authors 
\cite{Ali,Kagan,Deshpande1} that the decay rate of $B\to\eta' K$ will get
enhanced because of the small running strange quark mass at the scale $m_b$ 
and sizable $SU(3)$ breaking in the decay constants $f_8$ and $f_0$.
Ironically, it was also realized last year that \cite{Kagan,Ali} 
the above-mentioned enhancement is partially washed out by
the anomaly effect in the matrix element of pseudoscalar densities,
an effect overlooked before. Specifically, 
\be
\la\eta'|\bar s\gamma_5 s|0\ra=-i{m_{\eta'}^2\over 2m_s}\,\left(f_{\eta'}^s
-f^u_{\eta'}\right),
\en
where the QCD anomaly effect is manifested by the decay constant $f_{\eta'}
^u$. Since $f_{\eta'}^u\sim {1\over 2}f_{\eta'}^s$ 
it is obvious that the decay rate of $B\to\eta' K$ induced by the $(S-P)
(S+P)$ penguin interaction is suppressed by the
anomaly term in $\la\eta'|\bar s\gamma_5 s|0\ra$. As a consequence, the net
enhancement is not large. If we treat $\nc(LL)$ to be the same as $\nc(LR)$,
as assumed in previous studies, we would obtain typically 
${\cal B}(B^\pm\to\eta' K^\pm)=(2-3)\times 10^{-5}$ (see the dot-dashed curve
in Fig.~1).

  What is the role played by the intrinsic charm content of the $\eta'$ to
$B\to\eta' K$ ? It has been advocated that the new internal $W$-emission 
contribution 
coming from the Cabibbo-allowed process $b\to c\bar c s$ followed 
by a conversion of the $c\bar c$ pair into the $\eta'$ via two gluon 
exchanges is potentially important 
since its mixing angle $V_{cb}V_{cs}^*$ is as large as that
of the penguin amplitude and yet its Wilson coefficient $a_2$ 
is larger than that of penguin operators. 
The decay constant $f_{\eta'}^c$, 
defined by $\la 0|\bar c\gamma_\mu\gamma_5c|\eta'\ra=if_{\eta'}^c
q_\mu$, has been calculated theoretically \cite{Ali2,Araki}
and extracted phenomenologically from 
the data of $J/\psi\to\eta_c\gamma,\,J/\psi
\to\eta'\gamma$ and of the $\eta\gamma$ and $\eta'\gamma$ transition form
factors \cite{Ali,Petrov}; it lies in the range
--2.3 MeV $\leq f_{\eta'}^c\leq$ --18.4 MeV. The sign of $f_{\eta'}^c$ is 
crucial for the $\eta'$ charm content contribution. For a negative $f_{\eta'}
^c$, its contribution to $B\to\eta' K$ is constructive for $a_2>0$. 
Since $a_2$ depends strongly on $\nc(LL)$, we see that the 
$c\bar c\to\eta'$ mechanism contributes constructively
at $1/\nc(LL)>0.28$ where $a_2>0$,
whereas it contributes destructively at $1/\nc(LL)<0.28$ where $a_2$ becomes
negative. In order to explain the abnormally large branching ratio of 
$B\to\eta' K$, an enhancement from the $c\bar c\to\eta'$ mechanism is 
certainly welcome in order to improve the discrepancy between theory and 
experiment. This provides another strong support for $\nc(LL)\approx 2$.
If $\nc(LL)=\nc(LR)$, then ${\cal B}(B\to\eta' K)$ will be {\it suppressed}
at $1/\nc\leq 0.28$ and enhanced at $1/\nc>0.28$ (see the dashed curve
in Fig.~1 for $f_{\eta'}^c=-15$ MeV). If the preference for $\nc$ is 
$1/\nc\lsim 0.2$ (see e.g. \cite{Ali3}), then it is quite clear that
te contribution from the $\eta'$ charm content will make the
theoretical prediction even worse at small $1/\nc$ ! On the contrary, 
if $\nc(LL)\approx 2$, 
the $c\bar c$ admixture in the $\eta'$ will always lead to constructive 
interference irrespective of the value of $\nc(LR)$ (see the solid curve in 
Fig.~1).

\begin{figure}[ht]
\psfig{figure=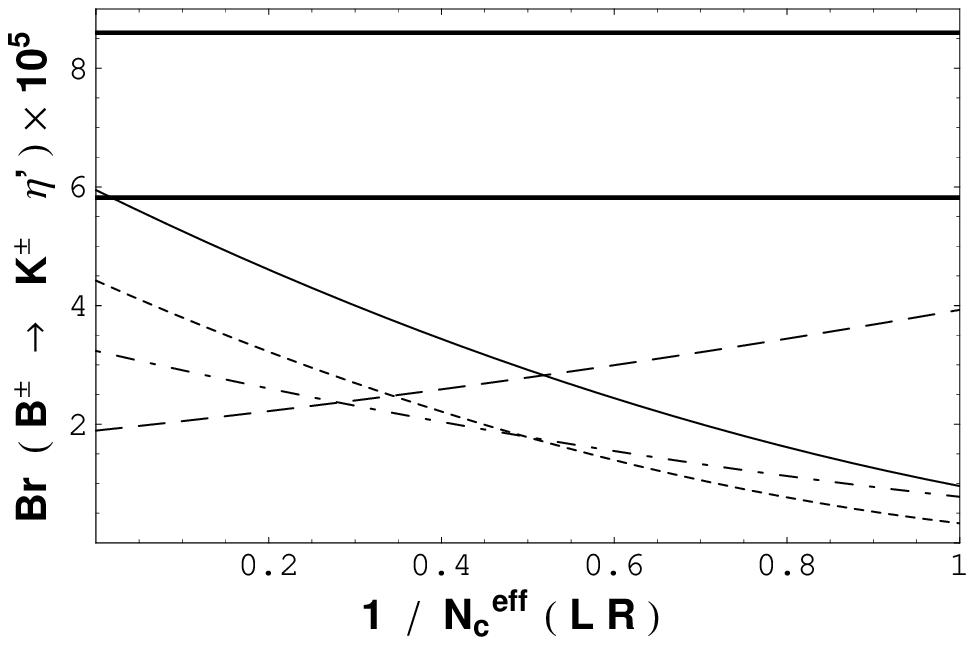,height=1.8in}
    \caption{{\small The branching ratio of $B^\pm\to\eta' K^\pm$ as a
   function of $1/\nc(LR)$ with $\nc(LL)$ being fixed at the value of 2
   and $\eta=0.34$, $\rho=0.16$.
   The charm content of the $\eta'$ with $f_{\eta'}^c=-15\,{\rm MeV}$ 
   contributes to the solid curve but not to the dotted curve.
   The anomaly contribution to $\la\eta'|\bar s\gamma_5s|0\ra$
   is included. For comparison, predictions for
   $\nc(LL)=\nc(LR)$ as depicted by the dot-dashed curve with $f_{\eta'}^c
   =0$ and dashed curve with $f_{\eta'}^c=-15$ MeV is also shown.
    The solid thick lines are the preliminary updated 
    CLEO measurements (13) with one sigma errors.}}
\end{figure}

   At this conference we learned that a recent CLEO reanalysis of 
$B\to\eta' K$  using a 
data sample 80\% larger than in previous studies yields the preliminary
results \cite{Behrens2,Roy}:
\be
{\cal B}(B^\pm\to\eta' K^\pm) &=& \left(7.4^{+0.8}_{-1.3}\pm 1.0\right)
\times 10^{-5},  \non \\
{\cal B}(B_d\to\eta' K^0) &=& \left(5.9^{+1.8}_{-1.6}\pm 0.9\right)
\times 10^{-5},
\en
suggesting that the original measurements (11) were not an upward statistical
fluctuation. This favors a slightly larger $f_{\eta'}^c$ in magnitude.
For $\nc(LL)=2$ and $f_{\eta'}^c=-15$ MeV, which is consistent with all the
known theoretical and phenomenological constraints, we show in Fig.~1 that
${\cal B}(B^\pm\to\eta' K^\pm)$ at $1/\nc(LR)\leq 0.2$ 
is enhanced considerably from $(2.5-3)\times 10^{-5}$ to $(4.6-5.9)
\times 10^{-5}$. In addition to the $\eta'$ charm content contribution,
$\nc(LL)\approx 2$ leads to constructive interference in the
spectator amplitudes of $B\to\eta' K$ and an enhancement in the term
proportional to $2(a_3-a_5)X_u^{(BK,\eta')}+(a_3+a_4-a_5)X_s^{(BK,\eta')}$.

  In the $B_s$ system, we \cite{CCT} find that $B_s\to\eta\eta'$, 
the analogue of $B^0\to\eta' K^0$, has also a large branching ratio of order
$2\times 10^{-5}$.

\subsection{Summary}
   Tree-dominated rare $B$ decays, $B^\pm\to\omega\pi^\pm$ and 
$B^0\to\pi^+\pi^-$, favor a small $\nc(LL)$, namely $\nc(LL)\approx 2$. The 
constraints on $\nc(LR)$ derived from the penguin-dominated decays
$B^\pm\to\phi K^\pm$ and $B\to\phi K^*$, which tend to be larger than 
$\nc(LL)$, are not consistent with each other. 
Our analysis of $B\to\eta' K$ 
clearly indicates that $\nc(LL)\approx 2$ is favored and $\nc(LR)$ is
preferred to be larger. The preliminary updated CLEO measurements of 
$B\to\eta' K$
seem to imply that the contribution from the $\eta'$ charm content is 
important and serious.

\section{Final-state interactions and $B\to\omega K$}
  The CLEO observation \cite{CLEOomega2} of a large branching ratio for 
$B^\pm\to \omega K^\pm$
\be
{\cal B}(B^\pm\to\omega K^\pm)=\left(1.5^{+0.7}_{-0.6}\pm 0.2\right)
\times 10^{-5},
\en
is difficult to explain at first sight. Its factorizable amplitude is
of the form
\be  \label{omegaK}
A(B^-\to\omega K^-) & \propto& (a_4+Ra_6)X^{(B\omega, K)}   \\
&+& (2a_3+2a_5+{1\over 2}a_9)X^{(BK,\omega)}+\cdots, \non
\en
with $R=-2m_K^2/(m_bm_s)$, where ellipses represent for contributions from 
$W$-annihilation and space-like penguin diagrams.
It is instructive to compare this decay mode closely with $B^-\to\rho K^-$ 
\be   \label{rhoK}
A(B^-\to\rho^0 K^-) \propto  (a_4+Ra_6)X^{(B\rho,K)}+\cdots.
\en
Due to the destructive interference between $a_4$ and $a_6$ penguin terms,
the branching ratio of $B^\pm\to\rho^0 K^\pm$ is estimated to be of order
$5\times 10^{-7}$. The question is then why is the observed rate of the
$\omega K$ mode much larger than the $\rho K$ mode ? By comparing 
(\ref{omegaK}) with (\ref{rhoK}),
it is natural to contemplate that the penguin contribution
proportional to $(2a_3+2a_5+{1\over 2}a_9)$ accounts for the large
enhancement of $B^\pm\to \omega K^\pm$. However, this is not the case: The
coefficients $a_3$ and $a_5$, whose magnitudes are smaller than $a_4$ 
and $a_6$, are not large enough to accommodate the data unless $\nc(LR)
<1.1$ or $\nc(LR)>20$ (see Fig.~9 of \cite{CT98}).

   So far we have neglected three effects in the consideration of 
$B^\pm\to\omega K^\pm$: $W$-annihilation, space-like
penguin diagrams and final-state interactions (FSI).
It turns out that FSI may play a dominant role for $B^\pm\to\omega K^\pm$.
The weak decays $B^-\to
K^{*-}\pi^0$ via the penguin process $b\to su\bar u$ and $B^-\to 
K^{*0}\pi^-$ via $b\to sd\bar d$ followed by the quark rescattering 
reactions $\{K^{*-}
\pi^0,~K^{*0}\pi^-\}\to\omega K^-$ contribute {\it constructively} to $B^-\to
\omega K^-$ (see Fig.~2), but {\it destructively} to $B^-\to\rho K^-$.
Since the branching ratios for $B^-\to K^{*-}\pi^0$ and
$K^{*0}\pi^-$ are large, of order $(0.5-0.8)\times 10^{-5}$, it is 
conceivable that a large branching ratio for $B^\pm\to\omega K^\pm$ can 
be achieved from FSI via inelastic scattering. Moreover, if FSI dominate,
it is expected that ${\cal B}(B^\pm\to\omega K^\pm)\approx (1+\sqrt{2})^2
{\cal B}(B^0\to\omega K^0)$.

\vskip 5mm
\begin{figure}[ht]
\psfig{figure=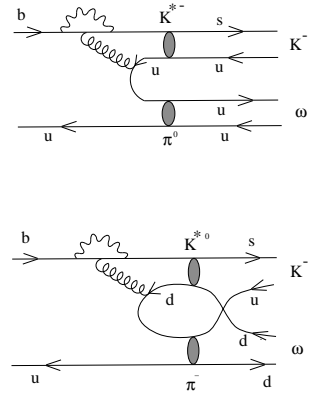,height=2.0in}
\caption{Contributions to $B^-\to K^-\omega$ from final-state interactions
via the weak decays $B^-\to K^{*-}\pi^0$ and $B^-\to K^{*0}\pi^-$ followed by
quark rescattering.}
\end{figure}

\section{Nonspectator Effects and $B$ Meson Lifetimes}

  In the heavy quark limit, all bottom hadrons have the same lifetimes,
a well-known result in the parton picture. With the advent of 
heavy quark effective theory and the
OPE approach for the analysis of inclusive weak decays, it is realized
that the first nonperturbative correction to bottom hadron lifetimes 
starts at order $1/m_b^2$ and it is model independent.
However, the $1/m_b^2$ corrections are small
and essentially canceled out in the lifetime ratios. 
The nonspectator effects such as $W$-exchange and
Pauli interference due to four-quark interactions are of order $1/m_Q^3$, 
but their contributions can be potentially significant due to a 
phase-space enhancement
by a factor of $16\pi^2$. As a result, the
lifetime differences of heavy hadrons come mainly from the above-mentioned 
nonspectator effects.

The four-quark operators relevant to inclusive nonleptonic $B$ decays are
\be\label{4qops}
   O_{V-A} &=& \bar b_L\gamma_\mu q_L\,\bar q_L\gamma^\mu b_L
    \,, \nonumber\\
   O_{S-P} &=& \bar b_R\,q_L\,\bar q_L\,b_R \,, \nonumber\\
   T_{V-A} &=& \bar b_L\gamma_\mu t^a q_L\,
\bar q_L\gamma^\mu  t^a b_L \,, \nonumber\\
   T_{S-P} &=& \bar b_R\,t^a q_L\,\bar q_L\, t^ab_R \,.
\en
From which one can follow \cite{NS} to define four hadronic parameters
$B_1,B_2,\varepsilon_1,\varepsilon_2$ relevant to our purposes:
\be \label{parameters}
{1\over 2m_{ B}}\la \ov B|O_{V-A(S-P)}|\ov B\ra &&\equiv {f^2_{B} m_{B}
\over 8}B_{1(2)}\,, \nonumber\\
{1\over 2m_{B}}\la \ov B|T_{V-A(S-P)}|\ov B\ra &&\equiv {f^2_{B} 
m_{B}\over 8}\varepsilon_{1(2)}\,.
\en
Under the factorization approximation, $B_i=1$ and $\varepsilon_i=0$.
To the order of $1/m_b^3$, the $B$-hadron lifetime ratios are 
given by
\begin{eqnarray}\label{ratios}
\frac{\tau(B^-)}{\tau(B^0_d)} &=& 1 +
   \Big( 0.043 B_1 + 0.0006 B_2
    - 0.61 \varepsilon_1 + 0.17 \varepsilon_2 \Big) \,, \nonumber \\
\frac{\tau (B^0_s)}{\tau(B^0_d)}
&=& 1+ 
(-1.7\times 10^{-5}\,B_1+1.9\times 10^{-5}\, B_2   \non \\
&& -0.0044 \varepsilon_1\, +0.0050\, \varepsilon_2)  \,.
  \end{eqnarray}
It is clear that
even a small deviation from the factorization approximation 
$\varepsilon_i=0$ can have a sizable impact on the lifetime ratios.

We have derived in heavy quark effective theory the renormalization-group 
improved QCD sum rules \cite{CK} for the hadronic parameters 
$B_1$, $B_2$, $\varepsilon_1$, and $\varepsilon_2$. The results are \cite{CK}
\be
B_1(m_b)=0.96\pm 0.04, &&  B_2(m_b)=0.95\pm 0.02, \non \\
\varepsilon_1(m_b)=-0.14\pm 0.01, &&  \varepsilon_2(m_b)=-0.08\pm 0.01,
\en
to the zeroth order in $1/m_b$.
The resultant $B$-meson lifetime ratios are $\tau(B^-)/\tau(B_d)=1.11\pm 
0.02$ and $\tau(B_s)/\tau(B_d)\approx 1$, to be compared with the world
averages \cite{Willoco}: $\tau(B^-)/\tau(B_d)=1.07\pm 
0.03$ and $\tau(B_s)/\tau(B_d)=0.94\pm 0.04$. Therefore, our prediction for
$\tau(B^-)/\tau(B_d)$ agrees with experiment.


\end{document}